\documentclass[aps,preprint,preprintnumbers,nofootinbib,showpacs,superscriptaddress]{revtex4-1}
\usepackage{graphicx,color,amsmath}
\begin{document}
\title{Semileptonic $B^-\to p\bar p \ell^- \bar\nu_{\ell}$ decays}

\author{C.Q. Geng}
\affiliation{Department of Physics, National Tsing Hua University, Hsinchu, Taiwan 300, R.O.C.}
\affiliation{Physics Division, National Center for Theoretical Sciences, Hsinchu, Taiwan 300, R.O.C.}
\author{Y.K. Hsiao}
\affiliation{Institute of Physics, Academia Sinica, Taipei, Taiwan 115, R.O.C.}
\date{\today}
\begin{abstract}
We study the four-body exclusive semileptonic baryonic $\bar B$ decays of $B^-\to p\bar p \ell^- \bar\nu_{\ell}$ ($\ell=e,\mu,\tau$)
 in the standard model. We find that their decay branching ratios
are about $(1.0, 1.0,0.5)\times 10^{-4}$, respectively.
In particular, the electron mode is close to the corresponding  CLEO's upper limit of  $5.2\times 10^{-3}$, while all results are about one or two orders of magnitude
larger than the previous estimated values for the
inclusive modes of $\bar B\to {\bf B\bar B'}\ell \bar \nu$.
Clearly, both B-factories of Belle and BaBar should be able to observe these exclusive four-body modes.
\end{abstract}


\maketitle
\newpage
\section{introduction}
In the semileptonic $\bar B\to M\ell\bar \nu_\ell$ decay with a meson $M$ and a charged lepton $\ell$,
the $\ell\bar \nu_\ell$ pair involves  no direct QCD interaction
so that the theoretical description of the amplitude can be reduced to a simple form  with the $\bar B\to M$ transition.
For example, the rate for  $\bar B^0\to \pi^+ e^-\bar \nu_e$  is proportional to $|V_{ub}f_+(q^2)|^2$,
where the form factor $f_+(q^2)$ for the $\bar B^0\to \pi^+$ transition depends on  the momentum transfer squared, $q^2$.
This benefits the precision measurement of $|V_{ub}|$,
where $|V_{ub}|$ is one of the least known Cabibbo-Kobayashi-Maskawa (CKM) matrix elements \cite{CKM1,CKM2} in the Standard Model (SM).
As long as we choose a point $q^2=q^2_i$ in the decay spectrum,
the corresponding data point with other parameters can be fixed to extract the value of $|V_{ub}|$.
However,
$f_+(q^2)$ relies on the calculations in the QCD models, such as
quark models \cite{quarkmodels}, lattice QCD \cite{latticeQCD}, and
Light Cone Sum Rules~\cite{lightcone}. Starting with $q^2_i$ and
$|V_{ub}|$, one is allowed to inversely extract the $q^2$ dependence
of $f_+(q^2)$ in different $q^2$ intervals from the measured
data~\cite{exf+_babar, exf+_belle, exF_cleo1, exF_cleo2, exF_babar}.
The extraction compared with various theoretical models hence
improves the knowledge of $f_+(q^2)$. Moreover, such extraction also
provides crosschecks for the $\bar B\to \rho$ and $\bar
B\to\eta^{(\prime)}$ transition form factors~\cite{exF_cleo1,
exF_cleo2, exF_babar}. In particular, the size of the gluonic
singlet contribution~\cite{singlet1,singlet2,singlet3} to the $\bar
B\to \eta^{\prime}$ transition to explain the unexpectedly large
two-body hadronic $\bar B\to K\eta^{\prime}$ decay rate has been
constrained by measuring $\bar B\to \eta^{(\prime)} \ell\bar\nu$
decays \cite{exF_cleo1, exF_cleo2, exF_babar}. Similar to the
mesonic cases, it should be interesting to extend the study to
baryonic decay modes, such as $\bar B\to {\bf B\bar B'}\ell\bar \nu$
with ${\bf B\bar B'}$ being a baryon pair, to investigate the $\bar
B\to {\bf B\bar B'}$ transition form factors, which have been used
as theoretical inputs in the three-body $\bar B\to p\bar p M$
decays.

\begin{figure}[t!]
\centering
\includegraphics[width=1.5in]{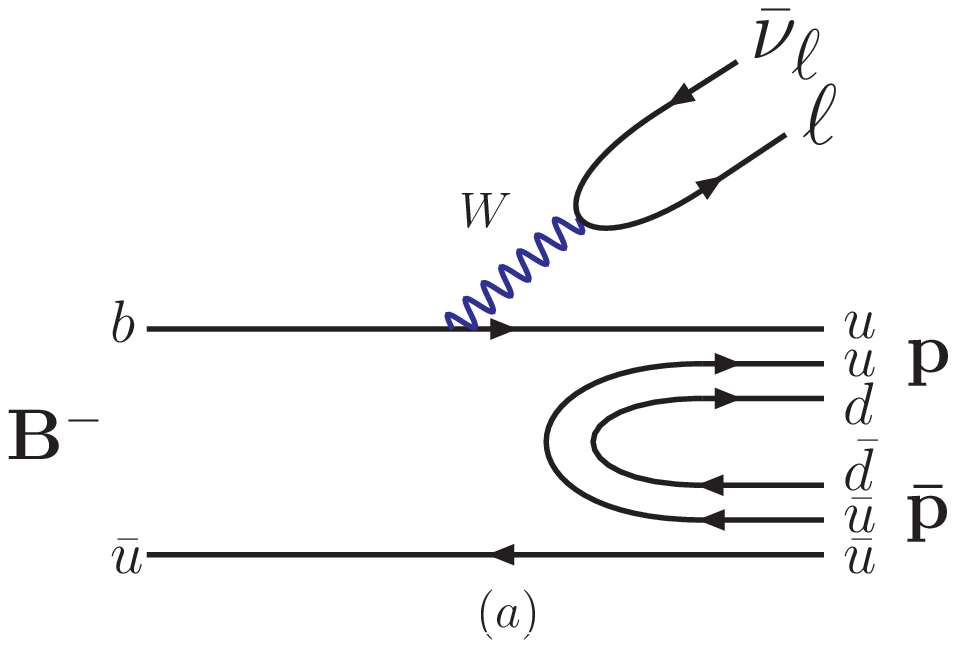}
\includegraphics[width=1.5in]{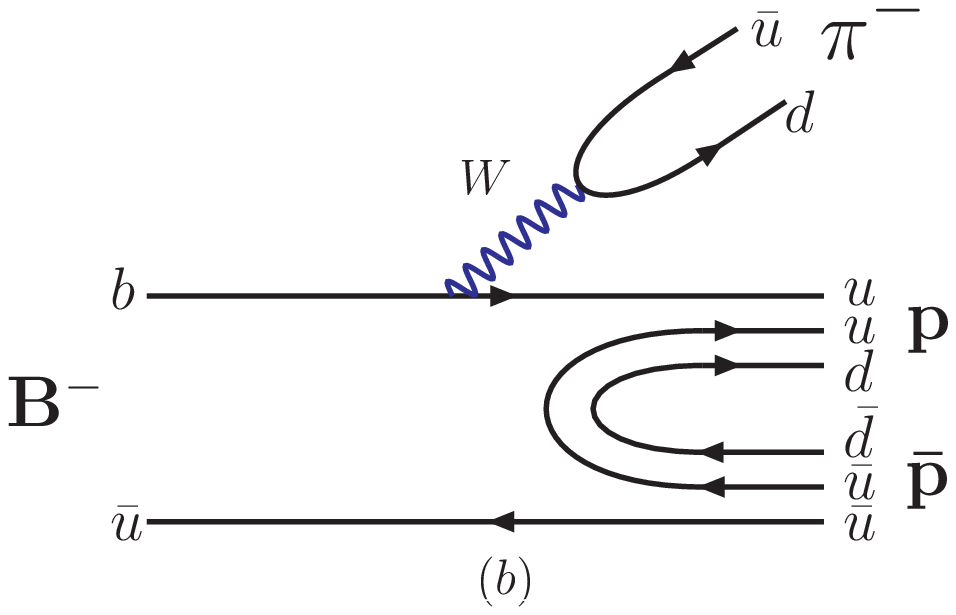}
\includegraphics[width=1.8in]{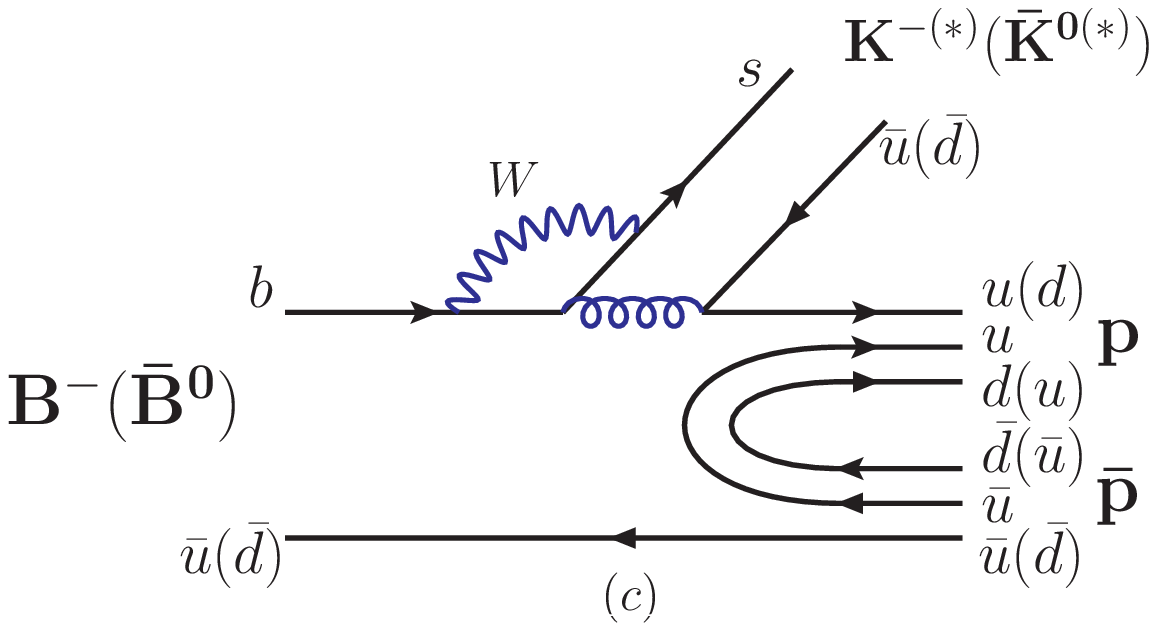}
\includegraphics[width=1.5in]{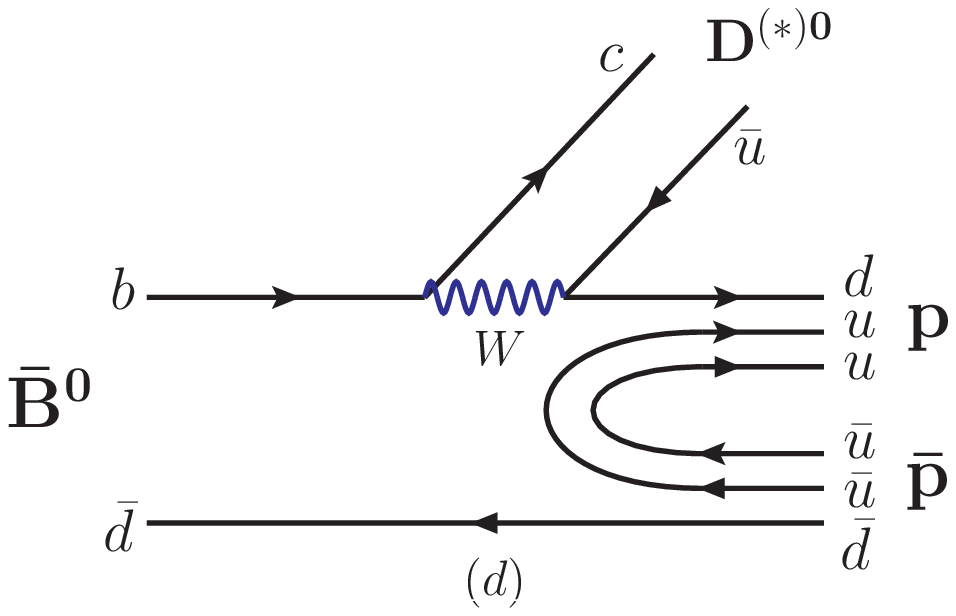}
\caption{Baryonic $\bar B$ decays with the $\bar B\to p\bar p$ transition, where
(a) $B^-\to p\bar p \ell\bar \nu_\ell$, (b) $B^-\to p\bar p \pi^-$,
(c) $\bar B\to p\bar p \bar K^{(*)}$, and (d) $\bar B^0\to p\bar p
D^{(*)0}$. }\label{fig1}
\end{figure}

The factorizable amplitudes for the three-body baryonic $\bar B\to {\bf B\bar B'}M$ decays are normally classified into current and
transition parts,
given by
\begin{eqnarray}
{\cal A_C}&\propto& \langle {\bf B\bar B'}|(\bar q_1 q_2)|0\rangle\langle M|(\bar q_3 b)|\bar B\rangle\,,\nonumber\\
{\cal A_T}&\propto& \langle M|(\bar q_1 q_2)|0\rangle\langle {\bf B\bar B'}|(\bar q_3 b)|\bar B\rangle\,,
\end{eqnarray}
respectively, where $(\bar q_1 q_2)$ and $(\bar q_3 b)$ stand for the weak currents. 
The matrix elements of $0\to {\bf B\bar B'}$
in $\cal A_C$ are presented as the timelike baryonic form factors, for which the theoretical calculations are available,
such as the approach of the pQCD counting rules \cite{Brodsky1,Brodsky2,Brodsky3}.
Consequently, the observed branching ratios for $\bar B^0\to n\bar pD^{*+}$ \cite{npbarD}, $B^-\to\Lambda \bar p\pi^-$
\cite{Lambdapbarpi1,Lambdapbarpi2,ppKLambdapbarpi_Belle,Lambdapbarpi3}
and $\bar B\to\Lambda \bar \Lambda \bar K^{(*)}$~\cite{LambdaLambdaK_Belle1,LambdaLambdaK_Belle2} can be explained
due to the
$\cal A_C$-like amplitudes~\cite{HY1,HY2,Hou1,Hou2,Hou3,GengHsiao1,GengHsiao2,GengHsiaoHY,Hsiao}.
On the other hand,
the measured decays of
$\bar B \to p\bar p \bar K^{(*)}$, $B^- \to p\bar p \pi^-$ \cite{ppKLambdapbarpi_Belle,ppK(star)pi_Belle,ppK_Babar,ppKpi_Belle,pppi_Babar,ppKstar_Belle},
and $\bar B^0\to p\bar p D^{(*)0}$ \cite{{ppD(star)_Belle,ppD(star)_Babar}}, shown in Fig.~ \ref{fig1},
are considered to have ${\cal A_T}$ as their amplitudes \cite{HY2,Hou2,GengHsiaoHY,Hsiao,GengHsiao3,GengHsiao4,GengHsiao5}.
To explain the data, the transition  matrix elements of $\bar B\to {\bf B\bar B'}$ are parameterized in terms of
various form factors
\cite{Hou2,GengHsiaoHY,Hsiao,GengHsiao3,GengHsiao4,GengHsiao5}.
As seen in Fig. \ref{fig1}, the decay of $B^-\to p\bar p  \ell \bar\nu_\ell$  is also  a ${\cal A_T}$ type as
 $\bar B \to p\bar p M$.
It is clear that
the observation of  $B^-\to p\bar p \ell \bar\nu_\ell$ shall
directly determine the transition form factors,
which have been widely used to explain the $\bar B\to p\bar p M$ data as theoretical inputs.
In analogy with the timelike baryonic form factors,
similar momentum dependences of  the transition form factors may be chosen,
which can be justified by investigating the shape of the invariant mass spectrum for the $B^-\to p\bar p \ell^- \bar\nu_\ell$ decay.
Moreover, we expect that the measurements of the angular distributions in $B^-\to p\bar p  e^- \bar\nu_e$
will provide some  information to understand the unexpectedly large angular distribution asymmetries of
${\cal A}_\theta(B^- \to p\bar p K^-)=0.45\simeq -{\cal A}_\theta(B^- \to p\bar p \pi^-)$ \cite{ppKpi_Belle}.

At present, the CLEO Collaboration has given  an experimental upper limit:~\cite{ppenu_cleo}
\begin{eqnarray}\label{limit1}
{\cal B}(B^-\to p\bar p e^- \bar \nu_e)<5.2\times 10^{-3}\;\text{(90\% C.L.)}\,,
\end{eqnarray}
while the theoretical estimation  has only been done for the inclusive $\bar B\to {\bf B\bar B'}\ell \bar \nu$
decays with charmless dibaryons, given by \cite{HouSoni}
\begin{eqnarray}\label{limit2}
{\cal B}(\bar B\to {\bf B\bar B'}\ell \bar \nu)\simeq 10^{-5}-10^{-6}\,.
\end{eqnarray}
In this paper, we concentrate on the exclusive four-body semileptonic baryonic decay
of $B^-\to p\bar p \ell \bar \nu_\ell$ ($\ell=e,\,\mu$, or $\tau$).
In particular, we will study its decay branching ratio in the SM.

The paper is organized as follows.
In  Sec. II, we provide the formalism, in which
we show the decay amplitude and rate of $B^-\to p\bar p \ell \bar \nu_\ell$ along with
the definitions of the transition form factors of  $\bar B\to {\bf B\bar B'}$.
We give our numerical results and discussions  in Sec. III.
In Sec. IV, we present the conclusions.

\section{Formalism}
In terms of the effective Hamiltonian, given by
\begin{eqnarray}
{\cal H}(b\to u \ell\bar \nu)&=&\frac{G_F V_{ub}}{\sqrt 2}\;\bar u\gamma_\mu (1-\gamma_5)b\; \bar \ell\gamma^\mu (1-\gamma_5)\nu\,,
\end{eqnarray}
for the $b\to u$ transition with the $W$ boson emission to $\ell \bar \nu$ at the quark level,
we easily factorize the amplitude for the $B^-\to p\bar p \ell \bar \nu_\ell$ decay to be
\begin{eqnarray}\label{amp}
{\cal A}(B^-\to p\bar p \ell \bar \nu_\ell)=\frac{G_F V_{ub}}{\sqrt 2}
\langle p\bar p|\bar u\gamma_\mu (1-\gamma_5)b|B^-\rangle \;\bar \ell\gamma^\mu (1-\gamma_5) \nu_\ell\;,
\end{eqnarray}
where we have parameterized the amplitude in terms of the transition matrix element of $\bar B\to p\bar p$.
With Lorentz invariance, the most general forms
 of  the $\bar B\to {\bf B\bar B'}$ transition form factors can be written as
\cite{GengHsiaoHY}
\begin{eqnarray}\label{transitionF}
\langle {\bf B}{\bf\bar B'}|\bar q'\gamma_\mu b|\bar B\rangle=
i\bar u(p_{\bf B})[  g_1\gamma_{\mu}+g_2i\sigma_{\mu\nu}p^\nu +g_3p_{\mu} +g_4(p_{\bf\bar B'}+p_{\bf B})_\mu +g_5(p_{\bf\bar B'}-p_{\bf B})_\mu]\gamma_5v(p_{\bf \bar B'}),~~
\nonumber\\
\langle {\bf B}{\bf\bar B'}|\bar q'\gamma_\mu\gamma_5 b|\bar B\rangle=
i\bar u(p_{\bf B})[ f_1\gamma_{\mu}+f_2i\sigma_{\mu\nu}p^\nu +f_3p_{\mu} +f_4(p_{\bf\bar B'}+p_{\bf B})_\mu +f_5(p_{\bf\bar B'}-p_{\bf B})_\mu]        v(p_{\bf \bar B'}),~~~
\end{eqnarray}
with $p=p_B-p_{\bf B}-p_{\bf\bar B'}$ for the vector and axial-vector quark currents, respectively. For the momentum dependences of $f_i$ and $g_i$,
we can rely on the results in the $\bar B\to p\bar pM$ decays as
they share the same $\bar B\to {\bf B\bar B'}$ transition form factors.
Since the $p\bar p$ invariant mass distributions for $\bar B\to p\bar pM$
have been observed to peak near the threshold area and flatten out at the large energy region,
inspired by the pQCD counting rules \cite{Brodsky1,Brodsky2,Brodsky3,Hou2}, we simply take the form factors as \cite{GengHsiao3}
\begin{eqnarray}\label{figi}
f_i=\frac{D_{f_i}}{t^n}\,,\;\;g_i=\frac{D_{g_i}}{t^n}\,,
\end{eqnarray}
with $n=3$ and $t\equiv (p_p+p_{\bar p})^2\equiv m_{p\bar p}^2$
where $D_{f_i}$ and $D_{g_i}$ are constants determined by the $\bar B\to p\bar p M$ data.
Note that the number of $n=3$ is for three hard gluons as the propagators to form a baryon pair in the approach of the pQCD counting rules,
where two of them attach to valence quarks in $p\bar p$, while the third one kicks and speeds up the spectator quark in $\bar B$.
In terms of Eqs. (\ref{amp}), (\ref{transitionF}), and (\ref{figi}),
the amplitude squared $|\bar {\cal A}|^2$ by summing over all fermion spins becomes available.

\begin{figure}[h!]
\centering
\includegraphics[width=2.5in]{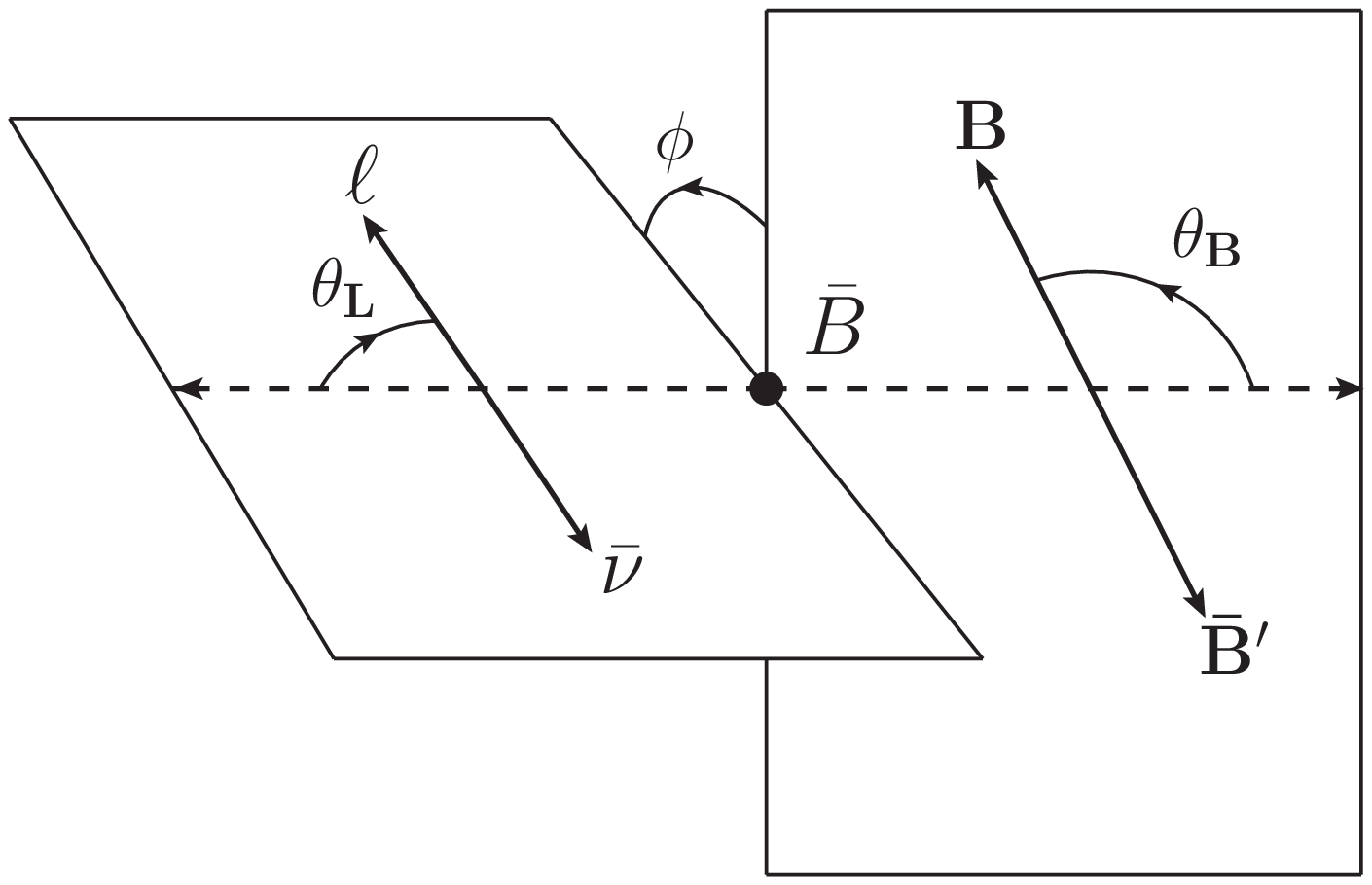}
\caption{Three angles of
$\theta_{\bf B}$, $\theta_{\bf L}$, and $\phi$
in $\bar B\to {\bf B}{\bf\bar B'}\ell\bar\nu$.}\label{3angles}
\end{figure}

We then need the kinematics for the four-body $\bar B(p_{\bar B})\to {\bf B}(p_{\bf B}) {\bf\bar B'}(p_{\bf\bar B'})\ell(p_\ell)\bar\nu (p_{\bar\nu})$ decay
to integrate over the phase space.
As the formalisms in $K_{l4}$, $D_{l4}$, and $B_{l4}$ decays
given in Refs. \cite{Kl4,Wise,Cheng:1993ah},
we use five kinematic variables,
$s\equiv (p_{\ell}+p_{\bar\nu})^2\equiv m_{\ell\bar \nu}^2$, $t$, $\theta_{\bf B}$, $\theta_{\bf L}$, and $\phi$
to describe the decay.
The three angles $\theta_{\bf B}$, $\theta_{\bf L}$, and $\phi$ are depicted in Fig. \ref{3angles},
where the angle $\theta_{\bf B(L)}$ is between $\vec{p}_{\bf B}$  ($\vec{p}_{\ell}$)
in the $\bf B\bar B'$ ($\ell\bar \nu$) rest frame and the line of flight of the $\bf B\bar B'$ ($\ell\bar \nu$) system in the rest frame
of the $\bar B$ meson,
while the angle $\phi$ is from the $\bf B\bar B'$ plane defined by the momenta of the $\bf B\bar B'$ pair
to the $\ell\bar \nu$ plane defined by the momenta of the $\ell\bar \nu$ pair in the rest frame of $\bar B$.
The partial decay width reads
\begin{eqnarray}
d\Gamma=\frac{|\bar {\cal A}|^2}{4(4\pi)^6 m_{\bar B}^3}X\beta_{\bf B}\beta_{\bf L}\, ds\, dt\, d\text{cos}\,\theta_{\bf B}\, d\text{cos}\,\theta_{\bf L}\, d\phi\,,
\end{eqnarray}
where $X$, $\beta_{\bf B}$, and $\beta_{\bf L}$ are given by
\begin{eqnarray}
X&=&\bigg[\frac{1}{4}(m_B^2-s-t)^2-st\bigg]^{1/2}\,,\nonumber\\
\beta_{\bf B}&=&\frac{1}{t}\lambda^{1/2}(t,m_{\bf B}^2,m_{\bf \bar B'}^2)\,,\nonumber\\
\beta_{\bf L}&=&\frac{1}{s}\lambda^{1/2}(s,m_{\ell}^2,m_{\bar \nu}^2)\,,
\end{eqnarray}
respectively,
with $\lambda(a,b,c)=a^2+b^2+c^2-2ab-2bc-2ca$.
The regions for the five variables of the phase space are given by
\begin{eqnarray}
&&(m_\ell+m_{\bar \nu})^2\leq s\leq (m_{\bar B}-\sqrt{t})^2\,,\;\;
(m_{\bf B}+m_{\bf \bar B'})^2\leq t\leq (m_{\bar B}-m_\ell-m_{\bar \nu})^2\,,\nonumber\\
&&0\leq \theta_{\bf L},\,\theta_{\bf B}\leq \pi\,,\;\;0\leq \phi\leq 2\pi\,.
\end{eqnarray}

\section{Numerical Results and Discussions }
\begin{figure}[h!]
\centering
\includegraphics[width=1.92in]{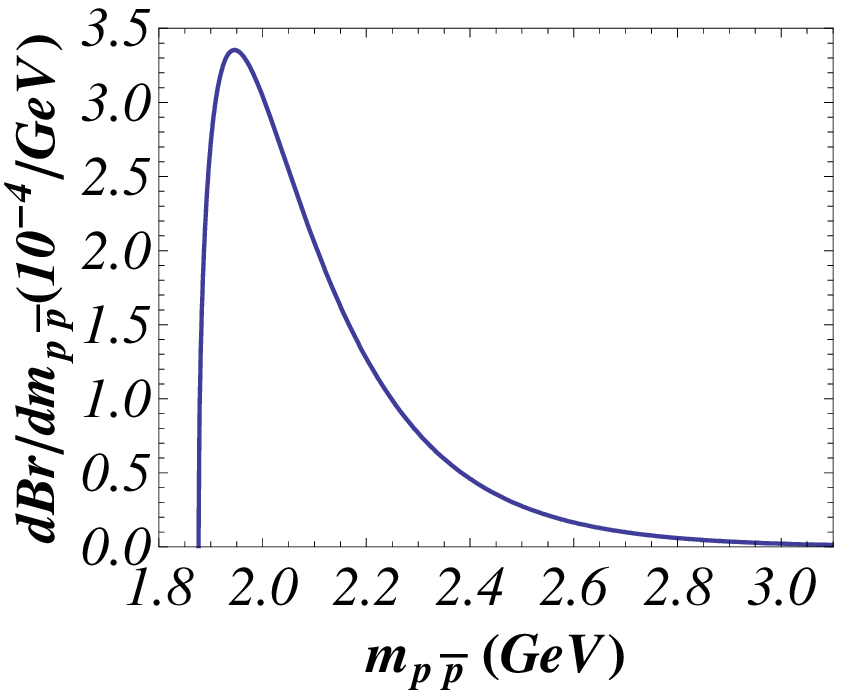}
\includegraphics[width=2in]{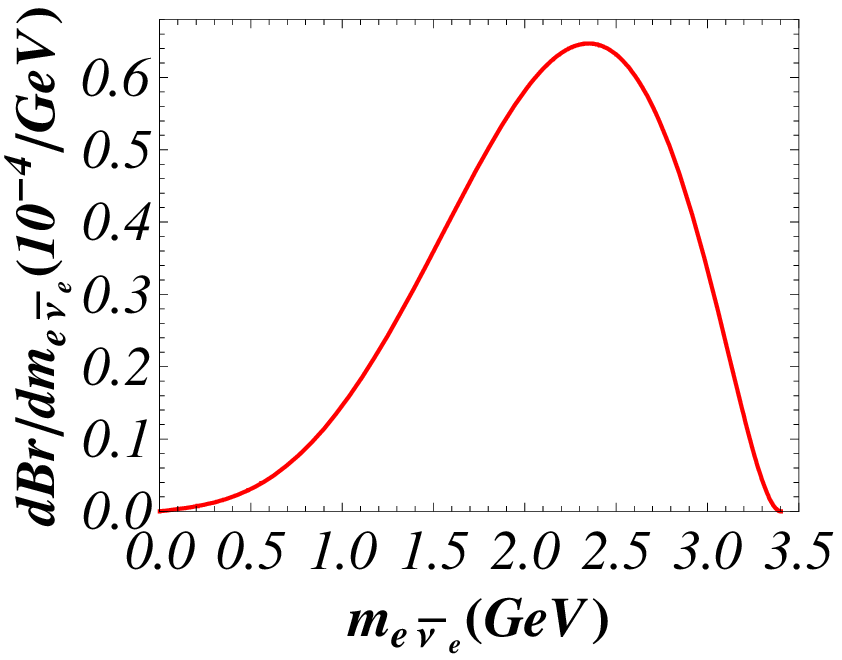}
\includegraphics[width=2in]{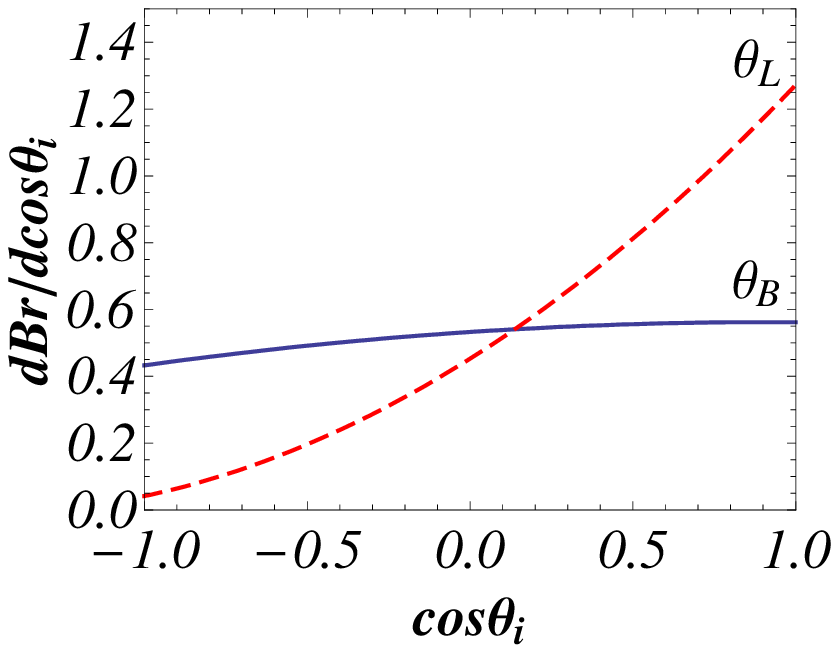}
\caption{Invariant mass spectra as functions of the invariant masses $m_{p\bar p}$ and $m_{e\bar \nu_e}$ and angular distributions
as functions of $\theta_i\ (i=B,L)$
for $B^-\to p\bar p e^-\bar \nu_e$, respectively.}\label{fig3}
\end{figure}

In our numerical analysis, we take $|V_{ub}|=(3.89\pm 0.44)\times 10^{-3}$ from the PDG \cite{pdg}.
To deal with $D_{g_i}$ and $D_{f_i}$ in Eq. (\ref{figi}), it is helpful to use the approach of the pQCD counting rules again,
where with  $SU(3)$ flavor and $SU(2)$ spin symmetries the vector and axial-vector currents are incorporated as two chiral currents
in the large t limit~\cite{Hou2,GengHsiaoHY,GengHsiao3}.
Consequently, $D_{g_i}$ and $D_{f_i}$ from the vector currents are related by another set of constants $D_{||}$ and  $D_{\overline{||}}$ from the chiral currents.
Explicitly, for the $B^-\to p\bar p$ transition form factors we have~\cite{GengHsiaoHY,GengHsiao3}
\begin{eqnarray}\label{D||}
D_{g_1}=\frac{5}{3}D_{||}-\frac{1}{3}D_{\overline{||}}\,,\;
D_{f_1}=\frac{5}{3}D_{||}+\frac{1}{3}D_{\overline{||}}\,,\;
D_{g_j}=\frac{5}{3}D_{||}^j=-D_{f_j}\,,
\end{eqnarray}
with $j=2,3,\, ..,\,5$, where their values are determined by fitting
the data of the total branching ratios, invariant mass spectra, and angular distributions
measured in the $\bar B\to p\bar p M$ decays.
To illustrate our results, we adopt the values in Ref. \cite{GengHsiaoHY}:
\begin{eqnarray}\label{inputD}
&&(D_{||},\;D_{\overline{||}})=(67.7\pm 16.3,\,-280.0\pm 35.9)\;{\rm GeV^5},\nonumber\\
&&(D_{||}^2,\,D_{||}^3,\,D_{||}^4,\,D_{||}^5)=\nonumber\\
&&(-187.3\pm 26.6,\,-840.1\pm 132.1,\,-10.1\pm 10.8,\,-157.0\pm 27.1)\;{\rm GeV^4}\;.
\end{eqnarray}
Thus, the total branching ratios of $B^-\to p\bar p \ell \bar\nu_\ell$ are found to be
\begin{eqnarray}\label{pred}
{\cal B}(B^-\to p\bar p e^- \bar\nu_e)&=&(1.04\pm 0.26\pm 0.12)\times 10^{-4}\,,\nonumber\\
{\cal B}(B^-\to p\bar p \mu^- \bar\nu_\mu)&=&(1.04\pm 0.24\pm 0.12)\times 10^{-4}\,,\nonumber\\
{\cal B}(B^-\to p\bar p \tau^- \bar\nu_\tau)&=&(0.46\pm 0.10\pm 0.05)\times 10^{-4}\,,
\end{eqnarray}
where the two errors in Eq. (\ref{pred}) are from those in Eq. (\ref{inputD}) and $|V_{ub}|$, respectively.
The invariant mass spectra and angular distributions for $B^-\to p\bar p e^- \bar\nu_e$
are shown in Fig. \ref{fig3}.
The integrated angular distribution asymmetries, defined by
\begin{eqnarray}
{\cal A}_{\theta_i}\equiv
\frac{\int^1_0 \frac{d{\cal B}}{dcos\theta_i}dcos\theta_i-\int^0_{-1}\frac{ d{\cal B}}{dcos\theta_i}dcos\theta_i}
{\int^1_0 \frac{d{\cal B}}{dcos\theta_i}dcos\theta_i+\int^0_{-1}\frac{ d{\cal B}}{dcos\theta_i}dcos\theta_i}\,,\ (i=B,\, L)
\label{AS}
\end{eqnarray}
are obtained to be
\begin{eqnarray}\label{angBL}
{\cal A}_{\theta_B}(B^-\to p\bar p e^- \bar\nu_e)&=&0.06\pm 0.02\,,\nonumber\\
{\cal A}_{\theta_L}(B^-\to p\bar p e^- \bar\nu_e)&=&0.59\pm 0.02\,,
\end{eqnarray}
where the errors are from those in Eq. (\ref{inputD}).

Since our result on ${\cal B}(B^-\to p\bar p e^- \bar\nu_e)$ in
Eq.~(\ref{pred}) is around $1.0\times 10^{-4}$, which is the same order of magnitude as those of the well measured
mesonic $B$ decays at Belle and BaBar,
such as ${\cal B}(\bar B^0 \to \pi^+(\rho^+)\ell^-\bar \nu_\ell)$ and ${\cal B}(B^- \to \rho^0\ell^-\bar \nu_\ell)$,
this four-body mode should be observed at these B-factories~\cite{MRWang}.
Moreover, as seen from Fig. \ref{fig3}a,
the $B^-\to p\bar p e^- \bar\nu_e$ decay inherits
the same threshold enhancement as those in the three-body baryonic $\bar B$ decays,
resulting from the adoption of  $1/t^3$ for the momentum dependence in the $B^-\to p\bar p$ transition form factors.
The spectrum in Fig. \ref{fig3}b reflects the fact that
in the helicity structure the amplitude of the $e^-\bar \nu_e$ pair is proportional to $(E_e+E_{\bar \nu_e})$.

It is interesting to note that our study of $B^-\to p\bar p \ell \bar\nu_\ell$ is similar to
that of $B^-\to p\bar p K^{*-}$~\cite{GengHsiao3,GengHsiao4,GengHsiao5}.
The terms related to $g_2$ and $f_2$ in the $B^-\to p\bar p$ transition form factors
give the main contributions to $B^-\to p\bar p \ell^- \bar\nu_{\ell}$.
Since
 the pair of the left-handed electron and the right-handed anti-neutrino in the helicity structure
behaves as one of the polarization vector $\varepsilon_{-}^\mu(p)$ with $p=p_{\ell}+p_{\bar \nu_{\ell}}$,
leading to $\varepsilon\cdot p=0$,
the contributions from $f_3$ and $g_3$ disappear.
Those from $f_4$ and $g_4$ are effectively small due to the tiny $|D_{||}^4|\simeq 10\,\text{GeV}^4$.
As the branching ratio receives the most contribution near the threshold area,
the $g_5$($f_5$)-accompanied term $(p_{\bar p}-p_{p})=(E_{\bar p}-E_p,\,\vec{p}_{\bar p}-\vec{p}_{p})\to (0,\,\vec{0})$ is suppressed.
Moreover, since the terms of $g_2$ and $f_2$ contain $\sigma_{\mu\nu}p^\nu$,
we have the relation
$|D_{g_2(f_2)}p|\simeq 300\,|p| \,\text{GeV}^5>$ $|D_{f_1}|\simeq 200\,\text{GeV}^5\gg$ $|D_{g_1}|\simeq 20\,\text{GeV}^5$,
which explains why  $g_2$ and $f_2$ prevail over other terms in the $B^-\to p\bar p \ell \bar\nu_\ell$ decay.

Finally, it is interesting to point out that
the angular distribution asymmetries in Eq.~(\ref{AS})
in the $B^-\to p\bar p \ell \bar\nu_\ell$ decay are sensitive to new physics,
such as the currents of
$(V+A)$ and $(S\pm P)$ beyond the SM.
Note that ${\cal A}_{\theta_L}=0.59$ in Eq. (\ref{angBL}) (see also Fig. \ref{fig3}c)
is caused by the $\ell^- \bar\nu_\ell$ pair of $(V-A)$ in the SM, which forms a polarization vector $\varepsilon_{-}^\mu(p)$
to couple to the left-handed helicity state of the virtual weak boson $W^{*-}$.
Therefore, a new physics with the $(V+A)$ current, which lets the $\ell\bar \nu_{\ell}$ pair to be another polarization state $\varepsilon_{+}^\mu(p)$,
must result in the deviation of  ${\cal A}_{\theta_L}$ in Eq. (\ref{angBL}).
Apart from a direct $CP$ violation~\cite{GengHsiao4}, 
$B^-\to p\bar p \ell \bar\nu_\ell$ can easily create $T$-odd triple product correlations (TPC's)
to test direct $T$ violation effects.
Since the three-momenta of $p\bar p$ and those of $\ell\bar \nu_\ell$ are not in the same plane,
$\vec{p}_\ell\cdot (\vec{p}_p\times \vec{p}_{\bar p})$ can be a nonzero TPC observable.
Like the case of $\vec{s}_\Lambda \cdot (\vec{p}_\Lambda\times \vec{p}_{\bar p})$ in $\bar B^0\to \Lambda\bar p \pi^-$~\cite{GengHsiao6}
with $\vec{s}_\Lambda$ denoting the $\Lambda$ spin,
there are other TPC observables $\vec{s}_\ell\cdot (\vec{p}_p\times \vec{p}_{\bar p})$ and $\vec{s}_p\cdot (\vec{p}_p\times \vec{p}_{\bar p})$ in $B^-\to p\bar p \ell \bar\nu_\ell$.
These rich TPC observables are expected to be useful to test new physics
in the advantage of ${\cal B}$ of order $10^{-4}$ much larger than the sensitivity of  $10^{-7}$ in the $B$ factories.

\section{Conclusions}
We have examined the four-body
 semileptonic baryonic $\bar B$ decay of  $B^-\to p\bar p \ell \bar\nu_\ell$ in the SM, which proceeds via
$b\to u \ell\bar \nu_\ell$ at the quark level.
The transition form factors  of $B^-\to p\bar p$,
which are well studied in the three-body baryonic $\bar B\to p\bar p M$ decays,
play the key role in the theoretical calculation.
We have found that  ${\cal B}(B^-\to p\bar p \ell \bar \nu_\ell)=(1.04,\,1.04,\,0.46)\times 10^{-4}$ for $\ell=e,\,\mu,\,\tau$, respectively,
which are just a little below the CLEO's upper limit of $5.2\times 10^{-3}$ for ${\cal B}(B^-\to p\bar p e^- \bar \nu_e)$ but much larger than
the previous estimations of $10^{-5}-10^{-6}$
 for the inclusive modes of $\bar B\to {\bf B\bar B'}\ell \bar \nu$.
 It is clear that the four-body decays of $B^-\to p\bar p \ell \bar\nu_\ell$, in particular the light charged leoton modes, should be observed by the B-factories of Belle and BaBar as well as
 future B-factories, such as Super-Belle and LHCb.

\begin{acknowledgments}
The work was supported in part by National Center of Theoretical Science
and  National Science Council
(NSC-98-2112-M-007-008-MY3)
of R.O.C.
\end{acknowledgments}

\end{document}